# How To Start a Grassroots Movement

David Ehrlich & Nora Szech*

September 2022

We study the influence of social messages that promote a digital public good, a COVID-19 tracing app. We vary whether subjects receive a digital message from another subject, and, if so, at what cost it came. Observed maximum willingness to invest in sending varies, from 1 cent up to 20 euros. Does this affect receivers' sending behavior? Willingness to invest in sending increases when previously receiving the message. Yet, cost signals have no impact. Thus, grassroots movements can be started at virtually no cost. App-support matters normatively as non-supporters are supposed to be punished in triage.

(*JEL* D9, I12, I18, H12)

(*Keywords* Grassroots movements, digital public good, social messages, COVID-19 tracing app, triage)

* Ehrlich: Corresponding author at Karlsruhe Institute of Technology (KIT), Department of Economics and Management, Kaiserstr. 12, 76131 Karlsruhe, Germany (email: david.ehrlich@kit.edu);
Szech: Corresponding author at Karlsruhe Institute of Technology (KIT), Department of Economics and Management, Kaiserstr. 12, 76131 Karlsruhe, Germany (email: nora.szech@kit.edu).
We thank Eugen Dimant, Egon Tripodi, Bernhard Kittel, Valerio Capraro, Frank Rosar, Ulrich von Ulmenstein, Christina von Lauppert von Peharnik, Max tretter, Matthias Braun, and Steffen Augsberg for helpful comments and suggestions. We thank Nicola Hüholt, Elena Fantino, and Paul Peters for excellent research assistance. This work was supported by the Federal Ministry of Education and Research of Germany (Grant number 01GP1905C).

Digital public goods become increasingly important (Goldfarb & Tucker 2019) and contributors typically face a trade-off between privacy costs and societal value. This applies to traffic apps warning how long the next congestion may last, disclosure of private health information to support research, as well as for COVID-19 tracing apps. Potentially, social messages could have strong impact on supporting such digital public goods. This is what our paper investigates. The recent global health crisis reemphasizes the significance of behavioral insights in spurring collective action (Bavel et al. 2020, Volpp et al. 2021), such as a widespread adoption of new technologies like tracing apps or vaccines. While digital contact tracing provides an effective tool to curb the spread of pandemics, uptake rates remain suboptimal. We study the effect of social messages (Capraro et al. 2021, Capraro et al. 2019, Dal Bó & Dal Bó 2014) and the effect of observed effort previously invested in sending.

We find that receiving a social message, encouraging tracing app use, significantly increases the willingness of receivers to invest in forwarding that message by around 50%. While subjects pay more for sending a message after receiving one, the effort provided by the previous sender has no significant impact on subjects' behavior. Even after observing that the sender was only willing to invest not more than 0.01 EUR for sending the social message, willingness to invest in receivers for passing on that message increases significantly, by almost 2.00 EUR, compared to the baseline where no message was received. Higher investments taken by the sender, even 20 euros, do not further increase the willingness to invest of receivers. This shows that grassroots movements can be started and entertained at very low costs, such as via social media.

With COVID-19 being the polarizing topic it is, the moral value of related health behavior is fiercely debated in the public sphere, even though health authorities issued recommendations on physical public goods like getting a vaccine, and on digital public goods such as using a tracing app. For instance, a Miami physician faced criticism for refusing to treat patients in person who chose to remain unvaccinated (Bella 2021). Therefore, this paper also employs an incentivized method (Krupka & Weber 2013) to elicit people's normative expectations concerning COVID-19-related and morally relevant public goods – one digital public good (tracing app), and one physical public good (vaccine). We also test the impact of social messages on these normative expectations. In particular, we examine norms on triage where subjects might or might not consider the patients' previous decisions regarding vaccination and support of tracing. Further, we examine the moral appropriateness of making patients pay (part of their) healthcare bills after being treated for COVID-19. Our results indicate that non-supportive behavior is deemed of highest moral relevance. In a situation as severe as triage, the prevailing norm we find is that non-supporters should be sacrificed.

People may invest more into sending a social message since it matters for their moral self-image (Benabou & Tirole 2002, Falk & Szech 2020, Loewenstein 1999). Or they may want to comply with a social norm (Bicchieri 2005, Bicchieri & Dimant 2019, Kittel et al. 2021, Schumpe et al. 2022). Receiving a social message could affect both. However, our data show that norms do not react to social messages. Thus, the increase in messaging support is likely driven by moral self-image.

We contribute in three ways. First, we find that receiving a social message, supporting a digital public good, from another person matters for willingness to invest in passing it on to others. Second, however, the investment taken by the sender for sending this message plays no significant role. This may explain why grassroots movements via social media work. Costs of sending are virtually nothing, yet this low-cost signal has no detrimental effect compared to much more costly price signals. Third, we study the effect of social messages on normative expectations and find that, in the highly debated topic of COVID-19, norms remain robust – suggesting that moral self-image concerns drive the increases in sending investments.

Digital contact tracing (Colizza et al. 2021) presents a useful instrument to fight a pandemic. By June 2021, more than 50 countries had introduced such exposure notification systems in response to the COVID-19 outbreak (O'Neill et al. 2020). Modeling studies and empirical evidence indicate their potential to curb the virus' spread (Wymant et al. 2021, Abueg et al. 2021, Kendall et al. 2020, Ferretti et al. 2020). Digital contact tracing can be effective even at low uptake rates (Moreno López et al. 2021). However, effectiveness increases substantially with larger user numbers (Aleta et al. 2020, Almagor & Picascia 2020, Kucharski et al. 2020). Thus, promoting contact tracing app use can contribute to public health, which is what many societies aim to achieve. Policymakers should be aware of the huge potential that grassroots movements may have here, at basically no cost.

We study nudges in the form of social messages (Thaler & Sunstein 2008). The literature contains a multitude of successfully implemented nudging interventions throughout a variety of themes and policy domains (Halpern 2015, Allcott 2011), including public health-related issues. Nudges may reduce missed hospital appointments (Hallsworth et al. 2015), encourage physical activity (Milkman et al. 2014), promote healthier dieting habits (Downs et al. 2009), or motivate people to get their flu vaccine shot (Milkman et al. 2021). Nudges can also help to spur organ donations (Johnson & Goldstein 2003) and blood donations (Goette & Tripodi 2020). There are studies employing nudges in the COVID-19 context. Serra-Garcia & Szech (2022) apply defaults to tackle vaccination hesitancy and increase willingness to get tested. An online study by Capraro & Barcelo (2020) examines the effect of messaging on intentions to wear a face mask. In contrast to our study, however, that study measures self-reported intentions, and the messages focus on risk perception rather than social information. In our study, behavior materializes.

While tracing app use yields public benefits, users face privacy costs when using the app (Grekousis & Liu 2021) and the app notifies users only after a potential exposure occured. There is even the risk of data security breaches. In Singapore, for example, authorities gave the police access to the collected data, who used it to conduct criminal investigations. The authorities had previously promised that they would use the data exclusively for COVID-19 contact tracing (Illmer 2021, O'Neill et al. 2020). Among other factors (Kaptchuk et al. 2020, Kozyreva et al. 2021, Rehse & Tremöhlen 2022), tracing app uptake, thus, depends on the willingness of people to disclose personal information (Montagni et al. 2021, Ross 2021, Benndorf et al. 2015, Beresford et al. 2012) and may also depend on their risk attitudes, which we elicit.

Needless to say, the COVID-19 pandemic will not be the last pandemic (Jones et al. 2008, Carlson et al. 2022), rendering what motivates or discourages people to use tracing apps even more relevant. Examples of other digital public goods abound. We consider understanding what drives people to support digital public goods (or not) to be as important as developing digital public goods in the first place. After all, what is the worth of a new technology if it is not supported? This is what our paper contributes to.

The remainder of this text is structured as follows: Section I lays out the experimental design. Section II presents our hypotheses, while section III provides the corresponding results. Section IV concludes.

## I.  Design

We conduct an incentivized online study with participants from the KD2lab subject pool at the Karlsruhe Institute of Technology in Karlsruhe, Germany. The participants previously signed up to take part in behavioral experiments. In total, we analyze the decisions of 709 individuals after excluding inconsistent – i.e., multi-switching in the multiple price list format described below – subjects. Our subjects are mostly university students with an average age of 24.56 years. Among the subjects, 59.5% identified as male. We implemented this study using the SoSci Survey tool and sent invitations to participate via KD2Lab's recruiting software hroot (Bock et al. 2014) and email. The invitation text informed subjects that a smartphone or mobile device would be required to participate. We ensured adherence by including a filter that only allowed mobile devices to access the online platform. Data collection took place from December 2020 to February 2021. At that time, the take-up rate in terms of active users of the governmentally recommended tracing app was estimated to be roughly between 25% and 35% of the 56 million smartphone users above the age of 15 in Germany. In terms of cumulative downloads up to that date, the share was approximately between 40% and 47% (Robert Koch-Institut 2022, Robert Koch-Institut/CWA Team 2022, Sueddeutsche Zeitung 2022).

At the top of the instructions, we informed subjects that, after the experiment, a computer program would randomly choose the subjects who would receive a payoff and whose decisions would be implemented precisely as stated in the instructions. Subjects knew one of their decisions would materialize with a probability of 10 percent. Since the study contains three incentivized parts, the computer would randomly pick one of them to be implemented.

Before making their individual decisions in the first part of the study, subjects would either – based on the randomized treatment assignment – receive one version of a social message, or no message at all. The participants who did not receive a message prior to their decisions constitute our control group. The content of the message that subjects in a treatment group received conveyed information about another person's willingness to invest in sending that message. The messages would read: "A randomly chosen person who also participated in this study, was willing to forego up to [monetary amount in EUR: 0.01; 2; 10; 20 – based on the randomized group assignment] in order to recommend using the COVID-19 Tracing App to you."

In the first part of the study, using the multiple price list format (Anderson et al. 2007), we elicited subjects' willingness to invest in sending the recommendation for using the COVID-19 tracing app. Specifically, subjects decided for each monetary amount whether they would prefer to forego that amount and send a message to another person or receive it as a payoff for themselves. Monetary amounts ranged from 0.01 EUR – intended to approximate the very low effort of sharing a message online via social media – up to 20 EUR. The latter mimics a situation in which passing on a message has high costs, which could be reputation costs, or cost and effort of delivering a message. Willingness to invest can also be taken as a signal as to how much subjects care about sending the message. Apparently, senders who only invest up to 1 cent for sending do not care much about the message.

The second incentivized part of the study elicited subjects' normative expectations for specific situations, using the methodology by Krupka & Weber (2013) to identify social norms. We presented a triage situation: Person A and Person B both need a ventilator due to falling ill with COVID-19. However, there is only one ventilator available. We asked participants whether they would deem it morally appropriate or inappropriate that Person A would be chosen for treatment, given two different circumstances. In the first case, Person A – contrary to Person B – had advocated against using the COVID-19 tracing app. In the second case, Person A – again contrary to Person B – had decided against getting a widely available and recommended vaccine. While supporting the tracing app is a digital public good, taking the vaccine has a physical public good character. We analogously applied both cases to a second situation, which addressed the costs to the public health sector caused by a COVID-19 treatment. Here, we asked if it would be morally appropriate for the statutory health insurance to recall part of the medical treatment cost from Person A. As the protocol by Krupka & Weber (2013) prescribes, the incentivization in this part of the study was designed such that subjects had an incentive to estimate the modal response, i.e., the prevailing social norm. Moral appropriateness was expressed on a 4-point-Likert scale with the corresponding ratings "morally very inappropriate, morally somewhat inappropriate, morally somewhat appropriate, morally very appropriate." As in Krupka & Weber (2013), we categorized responses as $-1$, $-1/3$, $1/3$, and $1$, accordingly, to analyze the results.

The third and final incentivized part of the study uses Holt & Laury's (2002) approach to elicit subjects' risk attitudes.

## II. Hypotheses

Our first two hypotheses concern the willingness to send a social message encouraging the use of a digital public good, in our case, a COVID-19 tracing app. We expect willingness to invest in sending this message to increase if participants received a social message themselves (Bond et al. 2012, Sisco & Weber 2019, Jordan et al. 2021).

**HYPOTHESIS 1.** *Receiving a social message supporting a digital public good increases the willingness to invest in passing on that message.*

We further expect that willingness to invest in sending the social message increases if participants know the sender was willing to invest more into sending it (Milgrom & Roberts 1986).

**HYPOTHESIS 2.** *Investment the sender was willing to make for sending the social message increases willingness to invest into passing on that message in the receiver.*

In addition, we are interested in social norms. The pandemic brought up various moral issues that were frequently and controversially discussed. One recurring problem was the risk of overburdening hospitals and other parts of the healthcare system. In the early days of the pandemic, exploding case numbers were out of control, e.g., in places like Northern Italy and New York City, leading to a shortage of ventilators. Doctors faced the grueling task of having to triage patients (Glenza 2020). Such triage problems remain relevant (Soltan et al. 2022, Connolly 2021) and have become a heavily discussed topic in many societies, including the United States (Schmidt et al. 2022). Therefore, we asked subjects – employing an incentive-compatible method (Krupka & Weber 2013) – about the social norm regarding a triage outcome between two patients. Subjects had the possibility to account for prior patient behavior. In particular, we studied two different cases. In the first case, one of the patients had previously discouraged others from using the COVID-19 tracing app (digital public good) before falling ill with COVID-19. In the second case, one of the patients rejected getting a vaccine (physical public good) before falling ill with COVID-19. In both cases, the other patients had not engaged in these non-supportive behaviors.

In addition to the triage problem, we examine a situation related to COVID-19 treatment costs and the potential tension between individual responsibility and common solidarity in the healthcare system. We asked subjects to evaluate the moral appropriateness of requiring patients to partly bear the cost of their COVID-19 treatment themselves (as opposed to having the insurance provider pay for it fully) after those patients previously opted against measures of public safety and caution before falling ill, reflected again in (a) opposing tracing app use, and (b) by refusing to get vaccinated against COVID-19.

As the literature shows, people are willing to punish – even if costly for themselves – the behavior of others acting in an anti-social or non-cooperative way (Fehr & Gächter 2002). We assume this to be applicable to digital public goods, like tracing apps, as well.

**HYPOTHESIS 3.** *Subjects consider it a norm to punish people who acted against a digital or physical public good. In the case of COVID-19, this reflects in norms on triage and treatment costs.*

Further, receiving a social message may underline the importance of the according norm (Milgrom & Roberts 1986).

**HYPOTHESIS 4.** *Social messages affect social norm perceptions in receivers.*

## III. Results & Discussion

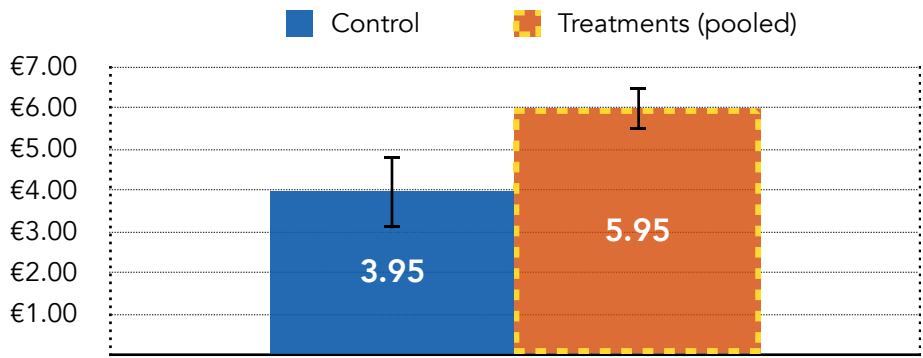

Figure 1. **Receiving a social message increases willingness to invest into sending**, substantially, in the receiver. Pairwise Wilcoxon rank-sum test: *p*=0.0001. Error bars indicate 95% confidence intervals.

### A. Willingness to invest

**RESULT 1.** *Receiving a social message on digital public good use increases willingness to invest in passing on that message.*

Receivers are willing to pay substantial amounts to pass on the social message supporting the use of the digital public good. When they do not receive a social message themselves (baseline), the willingness to invest averages 3.95 EUR (95% confidence interval = 3.07 EUR to 4.84 EUR). Turning to treatments groups where subjects received either the 1cent, 2EUR, 10EUR, or 20EUR social message, we find that subjects are willing to pay 5.95 EUR on average to pass on that social message (95% confidence interval = 5.43 EUR to 6.48 EUR). Compared to baseline, this amounts to a relative increase of roughly 50% (Figure 1).

All four treatments in which a social message was sent yield a positive effect on investment to pass on that message (p=0.0179 1 cent message, and p<0.01 respectively for the three other treatments) compared to baseline (Figure 2). Even when the observable maximum investment the sender was willing to make is only 1 cent, the messages are effective. Higher willingness to invest in sending does not translate into higher willingness to invest on the receiving side. While we find some variation in willingness to invest between the four treatments ranging from 5.48 EUR to 6.58 EUR, those differences are not statistically significant at any conventional level.

**RESULT 2.** *The observable maximum willingness to invest into sending the social message supporting the digital public good does not affect willingness to invest into passing it on.*

This shows that a grassroots movement can be easily started, even at virtually no costs. The finding underlines the power of social media, where costs of sending social messages are almost zero. Receivers become more motivated to invest in spreading a social message, even if they know the sender was only willing to invest 1 cent maximum into sending it.

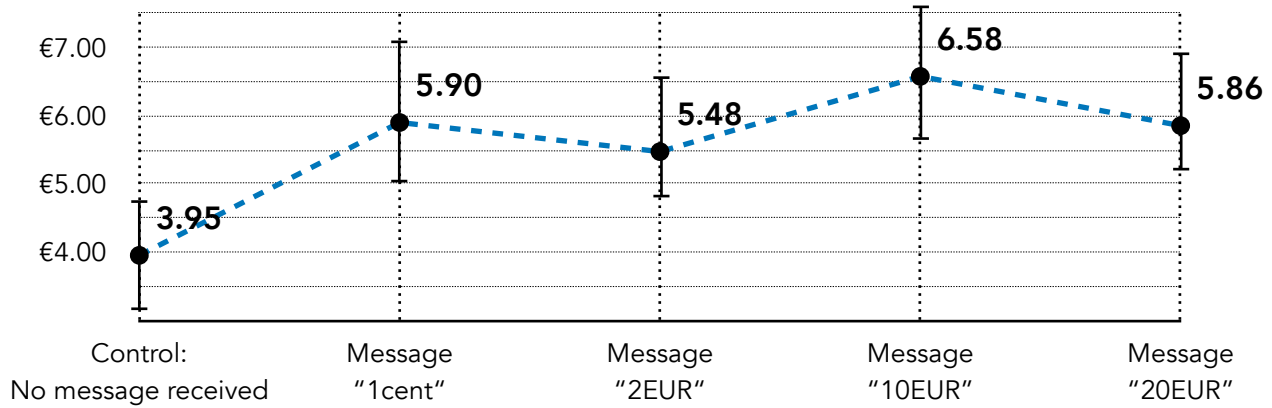

**Willingness to invest in sending, by treatments**

**Figure 2.** Receiving a social message affects messaging behavior of receivers. This figure presents the willingness to invest into sending a social message for the different treatment conditions. Stars indicate the p-values resulting from pairwise Wilcoxon rank-sum tests against baseline where no message was received prior to the decision (*=$p<0.1$, **=$p<0.05$, ***=$p<0.01$). **Willingness to invest does not differ significantly between the four message conditions.** Error bars indicate 95% confidence intervals.

Risk attitudes have no significant impact on willingness to invest (Figure 3). Possibly, participants with higher risk aversion place higher weight on privacy issues of the app (potentially decreasing willingness to invest), but also higher value on limiting the spread of the disease via app use (potentially increasing willingness to invest). Thus, effects may cancel out.

Moreover, as one might expect regarding previous observations of gender differences related to pro-social behavior (Eckel & Grossmann 1998), female subjects display substantially higher willingness to invest in sending (N=289, mean=€6,03 versus N=420, mean=€5,23; p=0.042, Wilcoxon rank-sum test). The interaction term between identifying as female and the treatment messages is, however, not significant based on a linear regression.

## B. Norms

The data indicate a social norm purporting that individuals who did not support a digital or physical public good should be sacrificed in a situation of triage (Figure 4). Thus, supporting both public goods is considered crucial in a life-or-death decision. Moreover, the social norm regarding vaccine behavior turns out to be very close to the social norm on app behavior – even though a causal link between tracing app use and eventually requiring life-saving treatment is much more difficult to substantiate in the case of not using the tracing app. By the same token, the proximity of both norms might stand out since the public debate in Germany surrounding tracing apps was controversial.

In Germany, paying directly for medical treatment is very uncommon, with insurances typically covering full treatment costs (Busse et al. 2017). Therefore, it may be difficult to find indications of a norm that non-supporters of a public good should pay part of their treatment. Indeed, the data

|  | (1) Investment in sending | (2) Norms: Triage/App | (3) Norms: Triage/Vax | (4) Norms: Cost/App | (5) Norms: Cost/Vax |
|---|---|---|---|---|---|
| Msg received? | 2.027 (0.591) | 0.0279 (0.0411) | -0.00313 (0.0478) | 0.0325 (0.0575) | 0.0276 (0.0620) |
| Gender=female | 0.896 (0.490) | 0.000312 (0.0340) | -0.0194 (0.0396) | 0.0182 (0.0476) | -0.0150 (0.0514) |
| Risk Attitude | -0.0575 (0.128) | -0.00622 (0.00888) | -0.000828 (0.0103) | -0.000242 (0.0124) | -0.0226 (0.0134) |
| Constant | 3.803 (0.842) | -0.322 (0.0585) | -0.365 (0.0681) | -0.330 (0.0819) | 0.0669 (0.0884) |
| Observations | 667 | 667 | 667 | 667 | 667 |
| Adjusted $R^2$ | 0.017 | -0.003 | -0.004 | -0.004 | 0.000 |
| F | 4.877 | 0.312 | 0.0874 | 0.150 | 1.083 |

Standard errors in parentheses

**Figure 3.** Social messages increase willingness to invest in sending, yet do not affect social norms. The table reports OLS estimates where the dependent variable is willingness to invest in sending (1), or the respective social norm(2)(3)(4)(5). Number of observations due to multi-switching in risk attitude elicitation.

indicate there is no social norm purporting that non-supporters should pay part of their treatment costs (Figure 4). Potentially, the norms in other countries might differ.

**RESULT 3.** *Subjects consider it a norm to punish non-supporters of the digital and of the physical public good when it comes to triage. In contrast, there is no norm that non-supporters should pay part of their treatment costs.*

Normative expectations appear remarkably constant across treatment variations (Figure 5). We find no evidence that receiving a social message affects norms on triage or treatment costs (Figure 3). Applying the Wilcoxon rank-sum test again for the different treatment groups individually against baseline, none of the social messages has a significant effect.

**RESULT 4.** *Receiving a social message does not affect norms regarding triage or treatment costs, neither in the digital public goods context (tracing app) nor in the physical public goods context (vaccine).*

Subjects may invest more into sending a social message as it is relevant to upholding their moral self-image (Benabou Tirole 2002, Falk & Szech 2020, Loewenstein 1999), or they might want to comply with a norm. Receiving social message might affect both. Our data show, however, that social messages do not influence norms in this context. Thus, the increase in sending behavior after receiving a social message is likely driven by moral self-image.

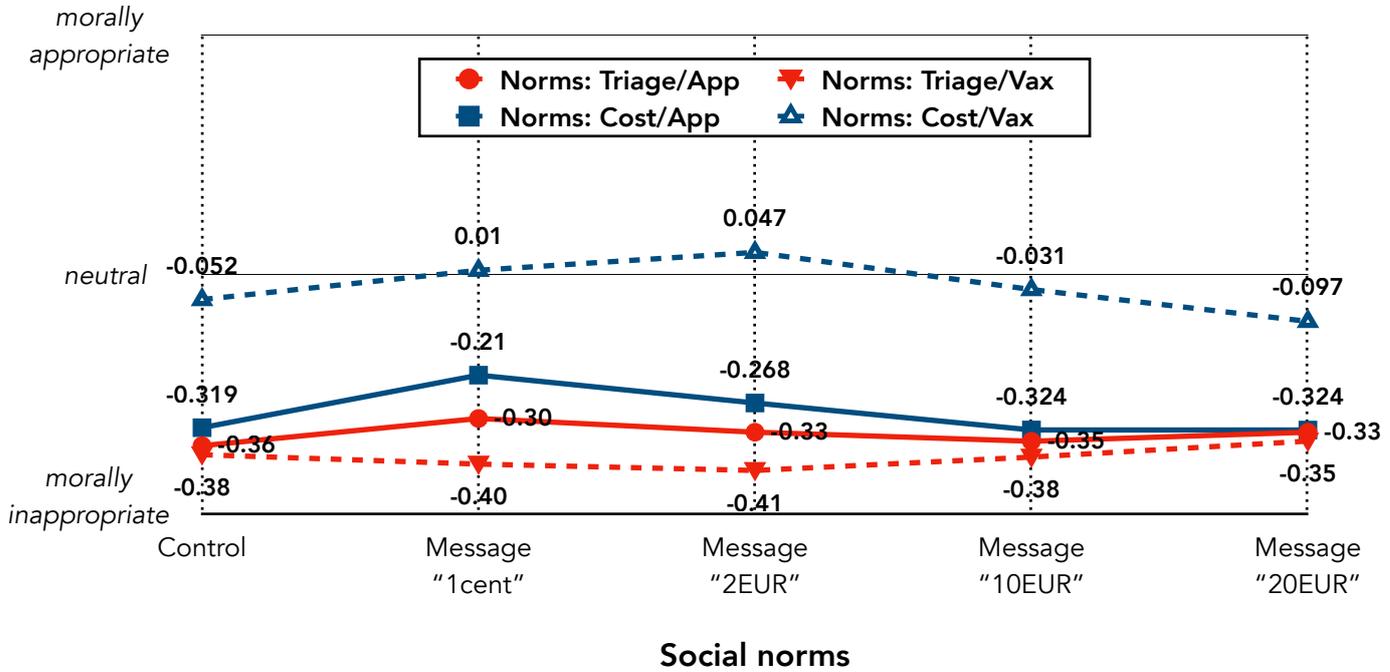

**Figure 4.** In a triage decision, subjects deem it morally inappropriate to prefer non-supporters over other patients. Burdening patients with treatment costs finds no normative support. Social messages have no significant impact. This figure shows subjects' average assessment of moral appropriateness. Following Krupka & Weber (2013), the response scale ranges from -1 = "very morally inappropriate" to 1 = "very morally appropriate".

## IV. Conclusion

It is often not salient how much effort people invest into sending a social message. Here, it is crystal clear to receivers how much senders invested. The data reveal that social messages work – no matter how large the cost a sender was willing to bear. Thus, we document that grassroots movements supporting a digital public good can be started at basically no cost.

The COVID-19 tracing app that our social messages were about, was supported by the government. Yet, public opinion about it was torn, in large part due to privacy concerns (Grekousis & Liu 2021, Miller & Solomon 2020). About 65–75 percent of the relevant population did not actively use the app when our studies took place. Our data document that even when a digital public good is controversial within society (Zimmermann et al. 2021, Meyer 2020), social messages can have an impact – independent of sending costs. Our data further document that using the tracing app mattered from a normative standpoint when it came to a situation as serious as triage. Thus, supporting or acting against digital public goods can bear highest moral relevance.

## Author contributions
DE and NS contributed equally to the research and article preparation. Both authors have approved the final version of this article.

## Conflict of interest
The authors declare no competing interests.

## Funding
This work was supported by the Federal Ministry of Education and Research of Germany (Grant number 01GP1905C).